# Exclusion limits on spin dependent WIMP-nucleon couplings from the SIMPLE experiment

F. Giuliani* and TA Girard

*Centro de Física Nuclear, Universidade de Lisboa, Av. Prof. Gama Pinto 2, 1649-003 Lisbon, Portugal*

**Abstract**

The SIMPLE experiment is reanalyzed within the context of the WIMP model independent framework of Tovey et. al.. The results are compared with those of the UK NAIAD and NaF experiments as recently reported by the Tokyo group, together with those of the latest UK NAIAD results in 2003, and further constrain the allowed parameter space.

The comparison of results from different spin-dependent WIMP searches conventionally [1] relies on a translation of WIMP-nucleus cross sections to WIMP-proton cross sections. This conversion requires assuming a value for the ratio of WIMP-proton and WIMP-neutron cross sections, which is highly WIMP-model dependent. Recently Tovey et. al. [2] have proposed an alternative method of analyses in which the customary conversion is replaced by a WIMP-model INdependent conversion to WIMP-nucleon cross sections. These considerations have been recently applied to interpretations of the UK Dark Matter [3], and the Tokyo LiF/NaF experiments [4]. A similar model-independent approach has been applied by DAMA [5].

Because of the negative sign in the ratio of proton-to-neutron spins in fluorine, the LiF/NaF measurements – although seemingly less significant than UKDM in the former exclusion phase space – have been shown to significantly reduce the allowed $\sigma_n$-$\sigma_p$ phase space of the UKDM experiment [4]. To complete the picture, we here provide a reanalyses of the previous SIMPLE experimental results [6] within the model-independent framework of Ref. 2. The results are compared with those of Ref. 4 and shown to further limit the parameter space. We additionally compare both F-based experiments with the most recent NAIAD results [3]. No comparison is made with DAMA, since their results for $^{23}$Na and $^{127}$I are not presented independently.

The procedure is well-described in Refs. 2. The general (zero momentum transfer) WIMP-nucleus

__________
* Corresponding author. Tel.: +351-21-790-4935; fax: +351-21-795-4228; e-mail: franck@cii.fc.ul.pt.

spin-dependent cross section $\sigma_A$ for a nucleus of mass number A is [1,2,7,8]:

$$\sigma_A = \frac{32}{\pi} G_F^2 \mu_A^2 \left( a_p \langle S_p \rangle + a_n \langle S_n \rangle \right)^2 \frac{J+1}{J} \quad , \qquad (1)$$

where $\langle S_{p,n} \rangle$ are the expectation values of the proton (neutron) group's spin, $a_{p,n}$ are the effective proton (neutron) couplings, $\mu_A$ is the WIMP-nuclide reduced mass, $G_F$ is the Fermi coupling constant, and J is the total nuclear spin.

If, as is usual in spin-dependent experiments, the search detector involves more than one nuclide, the limits on the allowed region in the $a_p$-$a_n$ plane become:

$$\sum_A \left( \frac{a_p}{\sqrt{\sigma_p^{\lim(A)}}} \pm \frac{a_n}{\sqrt{\sigma_n^{\lim(A)}}} \right)^2 < \frac{\pi}{24 G_F^2 \mu_p^2} \quad , \qquad (2)$$

where the sum extends over all nuclear species present in the detector's sensitive volume; $\mu_p$ is the WIMP-proton reduced mass, the mass difference between the two nucleons has been neglected, and $\sigma_p^{\lim(A)}$ ($\sigma_n^{\lim(A)}$) is the cross section limit when $a_n$ ($a_p$) = 0 respectively:

$$\sigma_p^{\lim(A)} = \frac{3}{4} \frac{J}{J+1} \frac{\mu_p^2}{\mu_A^2} \frac{\sigma_A}{\langle S_p \rangle^2} \qquad (3)$$

$$\sigma_n^{\lim(A)} = \frac{3}{4} \frac{J}{J+1} \frac{\mu_n^2}{\mu_A^2} \frac{\sigma_A}{\langle S_n \rangle^2} \quad .$$

The sign of the sum within the square of Eqn. (2) is that of the $\langle S_n \rangle / \langle S_p \rangle$ ratio.

Table 1 provides parameters used in the analyses. Since $^{37}$Cl has a magic number of neutrons (20), the odd group approximation has been assumed; it is the only species with $\sigma_n^{\lim(A)} = \infty$. For all other nuclei, published [2,4,7] spin values obtained from more refined calculations have been used. The listed abundances require multiplication by the respective detector composition fractions, which in the case of SIMPLE are 62.5 and 12.5% for F and Cl, respectively.

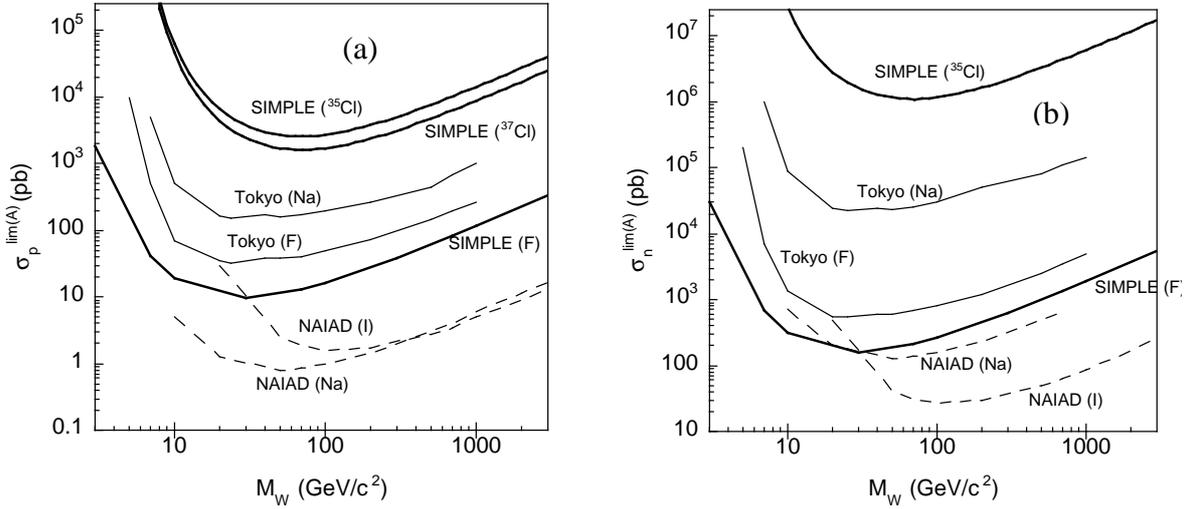

Fig. 1: (a) $\sigma_p^{\lim(A)}$ of SIMPLE (thick lines, 0.19 kg day), NAIAD (dashed lines, 3879 kg day), and Tokyo NaF (thin lines, 3.38 kg day) spin dependent WIMP-proton exclusion plots for each nuclide; (b) $\sigma_n^{\lim}$ of same.

Table 1: relevant spin values for the nuclides in this Letter. All are from Ref. 1, except $^{37}$Cl, calculated in the odd group model using data from Ref. 9.

| Nucleus | J | $<S_p>$ | $<S_n>$ | abundance (%) |
|---|---|---|---|---|
| $^{19}$F | ½ | 0.441 | -0.109 | 100 |
| $^{127}$I | 5/2 | 0.309 | 0.075 | 100 |
| $^{23}$Na | 3/2 | 0.248 | 0.020 | 100 |
| $^{35}$Cl | 3/2 | -0.083 | 0.004 | 76 |
| $^{37}$Cl | 3/2 | -0.178 | 0 | 24 |

Since $\sigma_{p,n}^{lim(A)}$ are functions of the WIMP mass $M_W$ only, Eqns. (2) and (3) are three parameter relations, which imply a 3D exclusion plot. As suggested in Ref. 2 and demonstrated in Ref. 4, we present $\sigma_{p,n}^{lim(A)}$ vs $M_W$, plus allowed regions in the $a_p$-$a_n$ plane, for selected $M_W$ values.

Fig. 1 shows the limit WIMP-proton and WIMP-neutron cross sections for all isotopes/experiments considered herein.

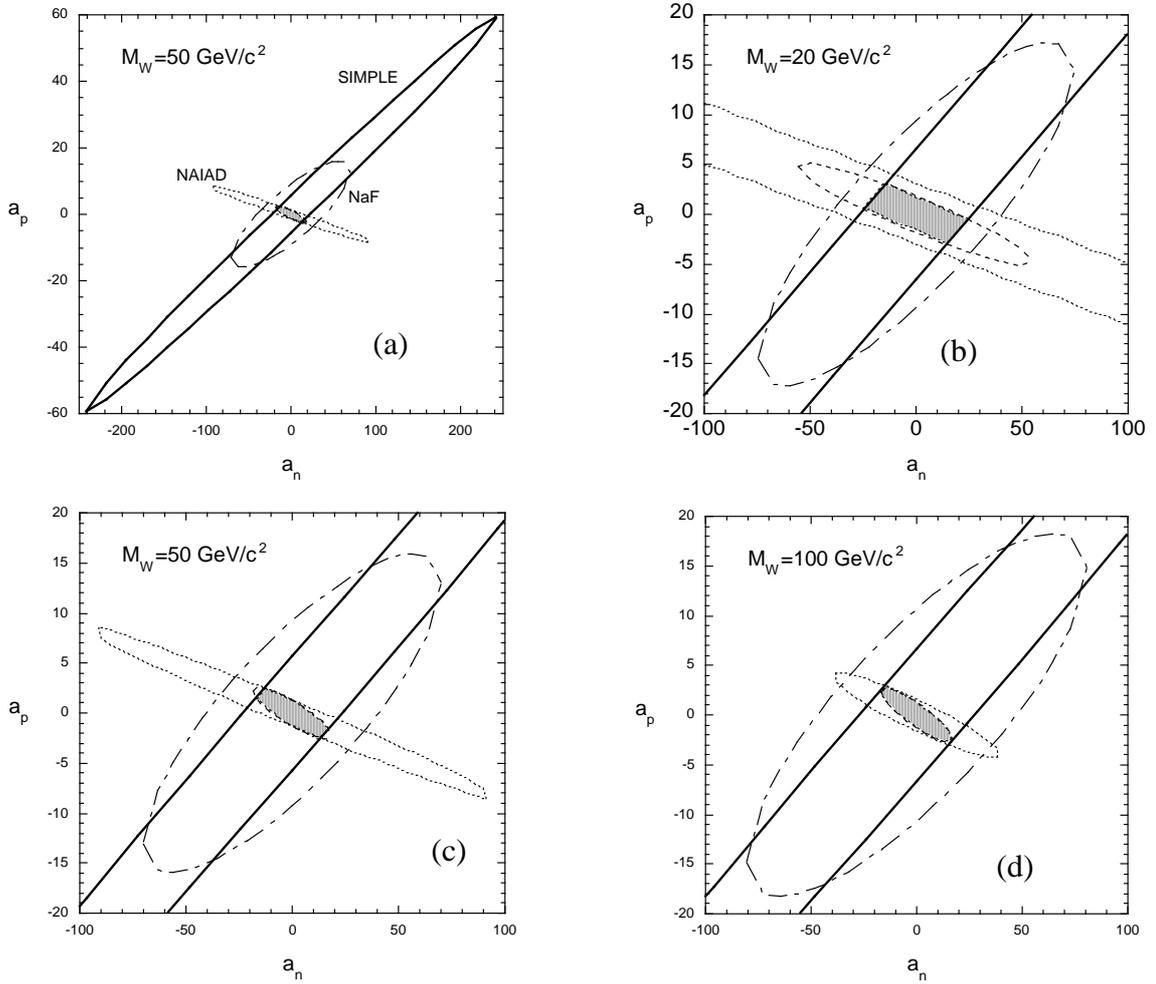

Fig. 2: spin dependent exclusion plots for WIMP mass of (a) 50 GeV/c$^2$, (b) 20, (c) 50, and (d) 100 GeV/c$^2$, with (a) and (d) identical except for scaling, for SIMPLE (solid), Tokyo NaF (dash-dotted); NAIAD results are represented as a dashed (2003 results) and a dotted (2000). The region permitted by each experiment is the area inside the respective ellipse.

Fig. 2 shows the reanalysis of 1999 SIMPLE results obtained with 0.19 kgdy exposure [6], together with the recent 3.38 kgdy NaF results by the Tokyo group [4], and two results by NAIAD: the 1080 kgdy used in Ref. 4 [10] and the improved 3879 kgdy published in 2003 [3]. The selected $M_W$ are 20 and 100 GeV/$c^2$, near those of optimal sensitivity for F and I respectively, and 50 GeV/$c^2$, which is in the DAMA-preferred range [11]. Fig. 2(a) shows the essential contour of the SIMPLE results, which is highly elliptical owing to the weaker constraint from Cl. Figs 2(b)-(d) show the selected $M_W$ in expanded scale, in which the SIMPLE ellipse appears as two parallel lines.

In all cases, both SIMPLE and the Tokyo experiments eliminate a large part of the phase space allowed by NAIAD 2000. In the case of $M_W$=50 GeV/$c^2$ for example, the SIMPLE results further reduce the parameter space allowed by the intersection of NAIAD and Tokyo by almost a factor two. When the improved NAIAD 2003 results are taken into account, which alone reduce the previously allowed parameter space by more than a factor three, the allowed regions reduce to the shaded areas of Fig. 2; for $M_W$=50 GeV/$c^2$, $|a_p| \leq 3$, $|a_n| \leq 17$.

The reason for the large impact of the fluorine-based detectors is (i) the relative sign of $<S_n>/<S_p>$ opposite to I, and (ii) both $<S_n>$ and $<S_p>$ non-negligible. The near-orthogonality of the fluorine ellipses results from (i); Cl and Na, the other nuclei present in the detectors, do not fulfill condition (ii), and are essentially spectators for WIMP-neutron interaction detection. Nevertheless, they make the unexcluded regions of the experiments closed ellipses instead of open conics [12], with the Cl being the weaker of the two constraint-wise. This weakness is responsible for the high eccentricity of the SIMPLE ellipse, owing to the low spin values of $^{35}$Cl and low concentration of $^{37}$Cl. Since SIMPLE's $\sigma_n^{\lim(F)}$ and $\sigma_p^{\lim(F)}$ are both one order of magnitude lower than the Tokyo group, both the NAIAD allowed regions in the $a_p$-$a_n$ plane are further reduced. A two order of magnitude improvement in results as projected by SIMPLE [13], if realized, would reduce the axes of the ellipse by ~ an order of magnitude yielding together with NAIAD 2003 $|a_p| \leq 1$, $|a_n| \leq 4$ for $M_W$=50 GeV/$c^2$.

**Acknowledgments**

We thank our colleagues in the SIMPLE collaboration for useful discussions. This work has been supported by POCTI/FNU/32493/2001 of the Portuguese Foundation for Science and Technology.